\documentclass[amsmath,amssymb,showkeys,superscriptaddress,floatfix,prd,10pt,aps,twocolumn]{revtex4-2}
\usepackage{graphicx,epstopdf}
\usepackage[caption=false]{subfig}
\usepackage{multirow}
\usepackage{appendix}
\def\pslash{p\!\!\!\slash }
\def\qslash{q\!\!\!\slash }
\def\xslash{x\!\!\!\slash }
\def\eslash{\varepsilon\!\!\!\slash }

\def\vel{\left|}
\def\ver{\right|}
\usepackage{colortbl}
\usepackage{xcolor}
\usepackage[colorlinks=true, urlcolor=blue, linkcolor=blue, citecolor=green]{hyperref}

\begin{document}

\title{Magnetic moment of the \texorpdfstring{$\Xi_b(6227)$}{} as a molecular pentaquark state}

\author{Ula\c{s} \"{O}zdem}%
\email[]{ulasozdem@aydin.edu.tr}
\affiliation{Health Services Vocational School of Higher Education, Istanbul Aydin University, Sefakoy-Kucukcekmece, 34295 Istanbul, Turkey}

\date{\today}
 
\begin{abstract}
In this study, considering that the $\Xi_b (6227)$ state is in molecular structure, the magnetic moment of this state is extracted in the light-cone QCD sum rules.
The numerical result is obtained as $\mu_{\Xi_b}=  0.12 \pm 0.03~\mu_N$.
The magnetic  moment of this state contains important information of its internal structure and shape deformations.
%
Measurement of the magnetic moment of the  $\Xi_b(6227)$ state in future experimental facilities can be very helpful in identification of the quantum numbers, as well as comprehension of the inner structure of this state.
\end{abstract}

\maketitle

\section{Introduction}\label{motivation}
In recent years, more and more heavy baryons have been observed by LHCb, Belle and CDF collaborations and the spectra of the heavy baryon sectors have become more and more abundance. Among the baryons, the singly bottom baryons are especially fascinating, because the heavy quark symmetry is preserved well in them, and their mass splitting is remarkably smaller than that in the case of light and charmed baryons.
Studies involving the spectroscopic parameters and decay channels of singly heavy baryons enhance our understanding of the non-perturbative regime of strong interaction, as well as our knowledge of their nature and internal structure.

In 2018, the LHCb Collaboration was reported a  bottom baryon, $\Xi_b (6227)$ ($\Xi_b$ for short), in both the $\Lambda^0_b K^-$ and $\Xi_b^0 \pi^-$ final states with  mass $m_{\Xi_b }= 6226.9 \pm 2.0 (\mbox{stat}) \pm 0.3 (\mbox{syst}) \pm 0.2~\mbox{MeV}  $ and width $\Gamma_{\Xi_b }= 18.1 \pm 5.4 (\mbox{stat}) \pm 1.8 (\mbox{syst})$~\cite{LHCb:2018vuc}. However, spin-parity of this state remains unknown. After the experimental discovery, many studies have been carried out to understand the properties of this particle. In some of these studies, this state is considered as the excited state of the $\Xi_b$ baryon \cite{Yu:2021zvl,He:2021xrh,Azizi:2020azq,Jia:2020vek,Nieves:2019jhp,Jia:2019bkr,Cui:2019dzj,Wang:2018fjm,Aliev:2018lcs,Chen:2018orb,Chen:2018vuc}, while in others it is considered as the pentaquark state~\cite{Wang:2020vwl,Zhu:2020lza, Huang:2018bed, Yu:2018yxl}.
Even though the studies of Refs.~\cite{Yu:2021zvl,He:2021xrh,Azizi:2020azq,Jia:2020vek,Nieves:2019jhp,Jia:2019bkr,Cui:2019dzj,Wang:2018fjm,Aliev:2018lcs,Chen:2018orb,Chen:2018vuc} seem to show that this state is a conventional three-quark state, the $\Xi_b$ might still be a $K \Sigma_b$ hadronic molecule state, because the mass gap between the $\Xi^*_b$ and the ground $\Xi_b$ baryons, about 440 MeV, is large enough to excite a light quark-anti-quark pair to form a molecular state.
In Ref.~\cite{Wang:2020vwl}, mass of the $\Xi_b$  state was studied in molecular pentaquark state with quantum numbers $J^P = 1/2^{\pm}$  via QCD sum rules. They found that their results support assigning $\Xi_b$ state to be a molecular state with quantum number $J^P = 1/2^{-}$. 
In Ref.~\cite{Zhu:2020lza}, the radiative decays of the $\Xi_b$ state have been obtained in the framework of effective Lagrangians by assuming this state as a molecular state with the quantum number $J^P =1/2^-$. 
In Ref.~\cite{Huang:2018bed}, the authors studied the strong decays of the $\Xi_b$ baryon assuming that it is a molecular state with different spin-party assignments. Their results showed that spin-parity  $J^P =1/2^-$ assignment is preferred while spin-parity  $J^P =1/2^+$ and $J^P =3/2^\pm$ are disfavored. In Ref.~\cite{Yu:2018yxl}, they studied mass of the $\Xi_b$ state in the molecular picture via an extension of the local hidden gauge approach in the Bethe–Salpeter equation. They found that the mass and width of $\Xi_b$ state remarkably close to the experimental data, for both the $J^P =1/2^-$ and $J^P =3/2^-$ sectors. However, the internal structure of this particle is still not clearly elucidated. In addition to spectroscopic properties such as mass and decay channels, different studies are needed to understand the properties of this state.

The investigation on the electromagnetic properties may provide a way of learning about the internal structure of the $\Xi_b$ state. Inspired by this, in the present study, we consider the magnetic moments of the $\Xi_b$ with the spin-parity  $J^P =1/2^-$, using light-cone QCD sum rule (LCSR) method ~\cite{Chernyak:1990ag, Braun:1988qv, Balitsky:1989ry} and assuming that the $\Xi_b$ is a hadronic molecular state. In the LCSR, the time-ordered product of the interpolating currents is sandwiched between the vacuum and an on-shell state. The on-shell state in the present work is the photon. As it is known, an appropriate correlation function is written while performing calculations in LCSR. Then, this correlation function is calculated in terms of both QCD degrees of freedom, QCD side, and hadronic parameters, hadronic side. The results obtained with these two different approaches are equalized with the dispersion relation by choosing the same Lorentz structures. Finally, Borel transform and continuum subtraction are performed to eliminate the contribution of excited and higher state terms.

This article is structured as follows. Section \ref{formalism} is set aside for the details of the LCSR evaluations for the magnetic moment of the $\Xi_b $ state. 
  Section \ref{numerical} is devoted to the numerical calculations of the magnetic moment and discussion.
 Some parameters and formulas related to the calculations are presented in Appendices A and B.
 \begin{widetext}

\section{Formalism}\label{formalism}

In this section, we explain how to compute the magnetic moment of $\Xi_b$ state in molecular picture. To derive the LCSR for the magnetic moment, we start by considering the subsequent correlation function,
%
 
\begin{eqnarray} \label{edmn01}
\Pi(p,q)&=&i\int d^4x e^{ip \cdot x} \langle0|T\left\{J^{\Xi_b}(x)\bar{J}^{\Xi_b}(0)\right\}|0\rangle _\gamma \, ,
\end{eqnarray}
where $\gamma$ is the external electromagnetic field and the $J^{\Xi_b}(x)$ stands for interpolating current of the considered state and it is given as follows:

\begin{align}
 J^{\Xi_b}(x) = [\bar u_d(x) i\gamma_5 s_d(x)][\varepsilon^{abc}u^T_a(x) C \gamma_\mu d_b(x)\gamma^\mu \gamma_5 b_c(x)], 
\end{align}
where $a,b,c~\mbox{and}~ d$ are color indices. It should be noted that the interpolating current couples to the five quark components of baryons. For a conventional baryon, this coupling would be small, but for a pentaquark state, this coupling can be large. In this work, we assume that $\Xi_b$ particle is dominantly a molecular state and hence  the coupling of the molecular interpolating current to $\Xi_b$ is not negligible.

As we mentioned above, the correlation function in the LCSR method is calculated in two different ways: hadronic and QCD sides. First, we focus on computation of the hadronic side. For this aim, we insert a full set of hadronic $\Xi_b$ state into Eq. (\ref{edmn01}) and carry out the integral over x to obtain
 \begin{align}\label{edmn02}
\Pi^{Had}(p,q)&=\frac{\langle0\mid J^{\Xi_b}(x) \mid
{{\Xi_b}}(p, s) \rangle}{[p^{2}-m_{{\Xi_b}}^{2}]}
\langle {{\Xi_b}}(p, s)\mid
{{\Xi_b}}(p+q, s)\rangle_\gamma 
\frac{\langle {{\Xi_b}}(p+q, s)\mid
\bar{J}^{\Xi_b}(0) \mid 0\rangle}{[(p+q)^{2}-m_{{{\Xi_b}}}^{2}]}+\cdots 
\end{align}

The matrix element, $\langle0\mid J^{\Xi_b}(x) \mid
{{\Xi_b}}(p, s) \rangle$ is written with respect to the residue and the polarization vector of the $\Xi_b$ state as
\begin{align}\label{res}
\langle0\mid J^{\Xi_b}(x) \mid
{{\Xi_b}}(p, s) \rangle = \lambda_{\Xi_b}u (p,s),
\end{align}
while the matrix element $\langle {{\Xi_b}}(p, s)\mid {{\Xi_b}}(p+q, s)\rangle_\gamma$ is written with respect to the form factors,
%
%
\begin{align} \label{edmn04}
\langle {{\Xi_b}}(p, s)\mid {{\Xi_b}}(p+q, s)\rangle_\gamma &=\varepsilon^\mu\,\bar u(p, s)\Big[\big[F_1(q^2)
+F_2(q^2)\big] \gamma_\mu +F_2(q^2)
\frac{(2p+q)_\mu}{2 m_{{\Xi_b}}}\Big]\,u(p+q, s).
\end{align}
%

We insert Eq.~(\ref{res}) and Eq.~(\ref{edmn04}) in Eq. (\ref{edmn02}), and summing over the polarization vector of the $\Xi_b$ state, we obtain the following result for the hadronic side of the correlation function:
\begin{align}
\label{edmn05}
\Pi^{Had}(p,q)=&\lambda^2_{{\Xi_b}}\gamma_5 \frac{\Big(\pslash+m_{{\Xi_b}} \Big)}{[p^{2}-m_{{{\Xi_b}}}^{2}]}\varepsilon^\mu \Bigg[\big[F_1(q^2) %
+F_2(q^2)\big] \gamma_\mu
+F_2(q^2)\, \frac{(2p+q)_\mu}{2 m_{{\Xi_b}}}\Bigg]  \gamma_5 
\frac{\Big(\pslash+\qslash+m_{{\Xi_b}}\Big)}{[(p+q)^{2}-m_{{{\Xi_b}}}^{2}]}. 
\end{align}
The value of  form factors $F_1(q^2)$ and $F_2(q^2)$ gives us the  magnetic form factor $F_M(q^2)$ at different $q^2$ :
\begin{align}
\label{edmn07}
&F_M(q^2) = F_1(q^2) + F_2(q^2).
\end{align}
At $q^2 = 0 $,  the magnetic moment $\mu_{{\Xi_b}}$ can be described with the $F_M (q^2 = 0)$ as follows
\begin{align}
\label{edmn08}
&\mu_{{\Xi_b}} = \frac{ e}{2\, m_{{\Xi_b}}} \,F_M(q^2 = 0).
\end{align}

Now, we turn our attention to calculate the QCD side of the above-mentioned correlation function. It is evaluated in deep Euclidean region where $p^2 \rightarrow \infty$ and $-(p + q)^2 \rightarrow \infty$. Employing the explicit expressions of the interpolating currents and contracting out the light and heavy quark pairs using the Wick’s theorem, we obtain

\begin{eqnarray}
\Pi^{QCD}(p,q)&=&-i\,\varepsilon_{abc}\varepsilon_{a^{\prime}b^{\prime}c^{\prime}}\, \int d^4x \, e^{ip\cdot x} \langle 0\mid  \gamma^\mu \gamma_5 S_{b}^{cc^\prime}(x) \gamma_5 \gamma^\nu  \nonumber\\
&&\mbox{Tr}\Big[\gamma_5 S_{u}^{dd^\prime}(-x) \gamma_5  S_{s}^{dd^\prime}(x)\Big]\, \mbox{Tr}\Big[\gamma_\mu S_d^{bb^\prime}(x)\gamma_\nu \widetilde S_{u}^{aa^\prime}(x)\Big]  \mid 0 \rangle _\gamma \, , 
\end{eqnarray}
where 
$\widetilde{S}_{c(q)}^{ij}(x)=CS_{c(q)}^{ij\mathrm{T}}(x)C$ and ,
 $S_{q}(x)$ and $S_{b}(x)$ are the light and b quark propagators, respectively. Their explicit expressions in the x-space are presented as
\begin{align}
\label{edmn13}
S_{q}(x)&= \frac{1}{2 \pi x^2}\Big(i \frac{\xslash}{x^2}- \frac{m_q}{2}\Big) 
- \frac{\langle \bar qq \rangle }{12} \Big(1-i\frac{m_{q} \xslash}{4}   \Big)
- \frac{ \langle \bar qq \rangle }{192}m_0^2 x^2  \Big(1
-i\frac{m_{q} \xslash}{6}   \Big)
-\frac {i g_s }{32 \pi^2 x^2} ~G^{\mu \nu} (x) \Big[\rlap/{x} 
\sigma_{\mu \nu} +  \sigma_{\mu \nu} \rlap/{x}
 \Big],\\
\nonumber\\
\label{edmn14}
S_{b}(x)&=\frac{m_{b}^{2}}{4 \pi^{2}} \Bigg[ \frac{K_{1}\Big(m_{b}\sqrt{-x^{2}}\Big) }{\sqrt{-x^{2}}}
+i\frac{{\xslash}~K_{2}\Big( m_{b}\sqrt{-x^{2}}\Big)}
{(\sqrt{-x^{2}})^{2}}\Bigg]
-\frac{g_{s}m_{b}}{16\pi ^{2}} \int_0^1 dv\, G^{\mu \nu }(vx)\Bigg[ (\sigma _{\mu \nu }{\xslash}
  +{\xslash}\sigma _{\mu \nu }) \frac{K_{1}\Big( m_{b}\sqrt{-x^{2}}\Big) }{\sqrt{-x^{2}}}
  \nonumber\\
  &
+2\sigma_{\mu \nu }K_{0}\Big( m_{c}\sqrt{-x^{2}}\Big)\Bigg],
\end{align}%
where $K_0$, $K_1$ and $K_2$ are modified the second kind Bessel functions and $G^{\mu\nu}$ is the gluon field strength tensor. 
The first term of the light and heavy quark propagators corresponds to perturbative or free part and the rest belong to the interacting parts. 
In the LCSR method, the non-perturbative contribution shows up when a photon is emitted at long distances. To take into account these contributions, it is required to expand the light quark propagator near the $ x^2 \sim 0$. In this case, the matrix elements of two and three-particle non-local operators show up between the photon states and vacuum such as $\langle \gamma(q)\vel \bar{q}(x) \Gamma_i q(0) \ver 0\rangle$ and $\langle \gamma(q)\vel \bar{q}(x) \Gamma_i G_{\mu\nu}q(0) \ver 0\rangle$. These matrix elements are parameterized in connection with the photon distribution amplitudes (DAs), which were determined in Ref. \cite{Ball:2002ps}. The explicit expressions of these terms are given in Appendix A.
The QCD side of the correlation function can be obtained associated with the quark-gluon properties via the photon DAs, and after performing an integration over x, the expression of the correlation function in the momentum representation can be calculated straightforwardly. 

As we explained in detail above, we have calculated the correlation function in two different ways. The next step will be to match this differently calculated correlation function with each other using certain approaches. To do this, we will use the quark-hadron duality approach and we  apply the double Borel transform and continuum subtraction to suppress the contribution of excited and higher states. Computations performed according to a scheme briefly explained above yield

\begin{align}
\label{edmn15}
\mu_{{\Xi_b}} \,\lambda^2_{{\Xi_b}}\, m_{{\Xi_b}} =e^{\frac{m^2_{{\Xi_b}}}{M^2}}\, \Delta^{QCD}.
\end{align}

The explicit expression of $\Delta^{QCD}$ is presented Appendix B.

\end{widetext}

\section{Numerical analysis and discussions}\label{numerical}

 The LCSR for magnetic moment of the ${\Xi_b}$ state contains many input parameters that we need their numerical values.  
We use $m_u=m_d=0$, $m_s =96^{+8}_{-4}\,\mbox{MeV}$, $m_b = 4.18^{+0.03}_{-0.02}\,\mbox{GeV}$~\cite{Patrignani:2016xqp}, $m_{\Xi_b}= 6226.9 \pm 2.0 (\mbox{stat}) \pm 0.3 (\mbox{syst}) \pm 0.2~\mbox{MeV}$~\cite{LHCb:2018vuc}, $\lambda_{\Xi_b} = (1.26^{+0.26}_{-0.27})\times 10^{-3} $~GeV$^6$~\cite{Wang:2020vwl},  $f_{3\gamma}=-0.0039~\mbox{GeV}^2$~\cite{Ball:2002ps}, 
$\langle \bar uu\rangle = 
\langle \bar dd\rangle=(-0.24 \pm 0.01)^3\,\mbox{GeV}^3$, $\langle \bar ss\rangle = 0.8\, \langle \bar uu\rangle$ $\,\mbox{GeV}^3$ \cite{Ioffe:2005ym},
$m_0^{2} = 0.8 \pm 0.1 \,\mbox{GeV}^2$ \cite{Ioffe:2005ym},  
$\langle g_s^2G^2\rangle = 0.88~ \mbox{GeV}^4$~\cite{Nielsen:2009uh} and $\chi=-2.85 \pm 0.5~\mbox{GeV}^{-2}$~\cite{Rohrwild:2007yt}. 
We also need the photon DAs and the input parameters used in these amplitudes to proceed in the calculations. The explicit expressions of the photon DAs and numerical values of input parameters are given in Appendix A.

In addition to these parameters mentioned above, QCD sum rules contain two extra parameters known as Borel mass parameter ($M^2$) and continuum threshold ($s_0$). According to the principles of the method used, the physical quantities to be calculated should not be very sensitive to the variation of these parameters. In practice, it is necessary to apply some physical constraints such as pole dominance and convergence of operator product expansion to ensure this situation. Considering these constraints, $50$~GeV$^2$ $ \leq s_0 \leq 52$~GeV$^2$ and $7$~GeV$^2$ $ \leq M^2 \leq 9$~GeV$^2$ working regions have been determined for these extra parameters. In Fig. 1, we  give the dependence of the magnetic moment of the $\Xi_b$ state on  Borel mass parameter $M^2$ , at three fixed values of the continuum threshold $s_0$. It follows from this figure that the magnetic moment indicates a weak dependence on $M^2$ in its working region.

 \begin{figure}[htp]
\centering
 \includegraphics[width=0.5\textwidth]{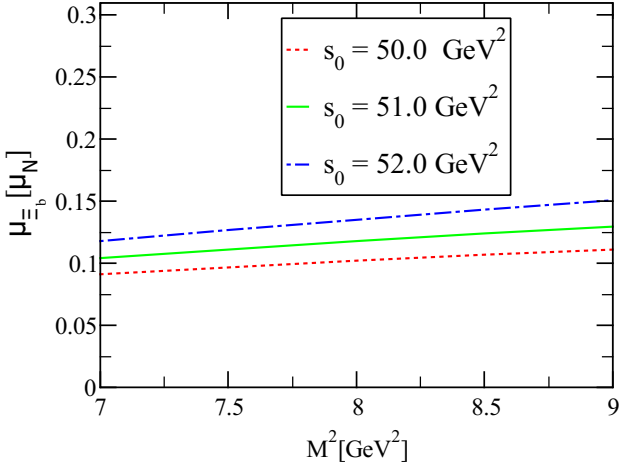} \caption{ The dependence of magnetic moment of the $\Xi_{b}$ state on $M^{2}$  at three fixed  values of $s_0$.}
  \end{figure}

We have specified all the input parameters necessary to determine the numerical value for the magnetic moment of the $\Xi_b$ state.
%
 %
 As a result of our detailed numerical calculations, where we consider the uncertainties in the input parameters, uncertainties entering the photon DAs as well as uncertainties because of the variations of Borel mass parameter $M^2$ and  continuum threshold $s_0$, the magnetic moment  of $\Xi_b$ state is finally found to have the value
\begin{align}
 \mu_{\Xi_b}= 0.80\pm 0.19~\frac{e}{2\,m_{\Xi_b}}= 0.12 \pm 0.03~\mu_N.
\end{align}
The order of magnetic moment shows that it is accessible in the experiment.  Detailed analysis of the calculations shows that the value of the magnetic moment is determined by the light quarks. 
Our analysis shows that the experimental measurement of the magnetic moment of the $\Xi_b$ will be able to tell whether it is a molecular state or a conventional baryon. If the measured magnetic moment value is consistent with our prediction, we can say that it is the molecular state; if not, we can say that it is the conventional baryon state. With the increased luminosity in future runs and our prediction, the facilities like  PANDA, LHCb, BESIII, Belle II and so on might have the capability to measure magnetic moment of $\Xi_b$ state.

Comparing the estimates obtained in this study with the results obtained using different approaches will be a test for the consistency of our results. The study of the magnetic moment of $\Xi_b$
with lattice QCD is strongly recommended.
In Refs.~\cite{Hall:2014uca,Hall:2016kou}, the mass and magnetic moment of $\Lambda(1405)$ state have been calculated  by considering it as the molecular picture in the framework of lattice QCD. 
A similar analysis can be performed using the Lattice QCD for the magnetic moment of the state $\Xi_b $. 
In comparison with the predicted magnetic moment of lattice QCD, the results in present study can ensure further information for the experimental search for the $\Xi_b$, and, moreover, the experimental measurements for this magnetic moment could be a important test for the molecule configuration of the $\Xi_b$. 
%

 
 In summary, inspired by experimental observation of the $\Xi_b(6227)$ state, the magnetic moment of the this state has been calculated in the framework of the light-cone QCD sum rules employing the photon DAs and assuming that the $\Xi_b(6227)$ is a hadronic molecular state. 
 The acquired result may be useful in exact determinations of the nature of this state.
 %
Checking our prediction with different theoretical models and  by future experiments can be very helpful of the comprehension of the inner structure as well as the geometric shape of the $\Xi_b(6227)$ state. 

 \begin{widetext}
 
 \appendix
 \section*{Appendix A: Photon distribution amplitudes and wave functions}

In this Appendix, we give the matrix elements $\langle \gamma(q)\vel \bar{q}(x) \Gamma_i q(0) \ver 0\rangle$  
and $\langle \gamma(q)\vel \bar{q}(x) \Gamma_i G_{\mu\nu}q(0) \ver 0\rangle$ in connection with  the photon DAs  and wave functions of different  twists \cite{Ball:2002ps},

\begin{eqnarray*}
\label{esbs14}
&&\langle \gamma(q) \vert  \bar q(x) \gamma_\mu q(0) \vert 0 \rangle
= e_q f_{3 \gamma} \left(\varepsilon_\mu - q_\mu \frac{\varepsilon
x}{q x} \right) \int_0^1 du e^{i \bar u q x} \psi^v(u)
\nonumber \\
&&\langle \gamma(q) \vert \bar q(x) \gamma_\mu \gamma_5 q(0) \vert 0
\rangle  = - \frac{1}{4} e_q f_{3 \gamma} \epsilon_{\mu \nu \alpha
\beta } \varepsilon^\nu q^\alpha x^\beta \int_0^1 du e^{i \bar u q
x} \psi^a(u)
\nonumber \\
&&\langle \gamma(q) \vert  \bar q(x) \sigma_{\mu \nu} q(0) \vert  0
\rangle  = -i e_q \langle \bar q q \rangle (\varepsilon_\mu q_\nu - \varepsilon_\nu
q_\mu) \int_0^1 du e^{i \bar u qx} \left(\chi \varphi_\gamma(u) +
\frac{x^2}{16} \mathbb{A}  (u) \right) \nonumber \\ 
&&-\frac{i}{2(qx)}  e_q \bar qq \left[x_\nu \left(\varepsilon_\mu - q_\mu
\frac{\varepsilon x}{qx}\right) - x_\mu \left(\varepsilon_\nu -
q_\nu \frac{\varepsilon x}{q x}\right) \right] \int_0^1 du e^{i \bar
u q x} h_\gamma(u)
\nonumber \\
&&\langle \gamma(q) | \bar q(x) g_s G_{\mu \nu} (v x) q(0) \vert 0
\rangle = -i e_q \langle \bar q q \rangle \left(\varepsilon_\mu q_\nu - \varepsilon_\nu
q_\mu \right) \int {\cal D}\alpha_i e^{i (\alpha_{\bar q} + v
\alpha_g) q x} {\cal S}(\alpha_i)
\nonumber \\
&&\langle \gamma(q) | \bar q(x) g_s \tilde G_{\mu \nu}(v
x) i \gamma_5  q(0) \vert 0 \rangle = -i e_q \langle \bar q q \rangle \left(\varepsilon_\mu q_\nu -
\varepsilon_\nu q_\mu \right) \int {\cal D}\alpha_i e^{i
(\alpha_{\bar q} + v \alpha_g) q x} \tilde {\cal S}(\alpha_i)
\nonumber \\
&&\langle \gamma(q) \vert \bar q(x) g_s \tilde G_{\mu \nu}(v x)
\gamma_\alpha \gamma_5 q(0) \vert 0 \rangle = e_q f_{3 \gamma}
q_\alpha (\varepsilon_\mu q_\nu - \varepsilon_\nu q_\mu) \int {\cal
D}\alpha_i e^{i (\alpha_{\bar q} + v \alpha_g) q x} {\cal
A}(\alpha_i)
\nonumber \\
&&\langle \gamma(q) \vert \bar q(x) g_s G_{\mu \nu}(v x) i
\gamma_\alpha q(0) \vert 0 \rangle = e_q f_{3 \gamma} q_\alpha
(\varepsilon_\mu q_\nu - \varepsilon_\nu q_\mu) \int {\cal
D}\alpha_i e^{i (\alpha_{\bar q} + v \alpha_g) q x} {\cal
V}(\alpha_i) \nonumber
\end{eqnarray*}
\begin{eqnarray*}
&& \langle \gamma(q) \vert \bar q(x)
\sigma_{\alpha \beta} g_s G_{\mu \nu}(v x) q(0) \vert 0 \rangle  =
e_q \langle \bar q q \rangle \left\{
        \left[\left(\varepsilon_\mu - q_\mu \frac{\varepsilon x}{q x}\right)\left(g_{\alpha \nu} -
        \frac{1}{qx} (q_\alpha x_\nu + q_\nu x_\alpha)\right) \right. \right. q_\beta
\nonumber \\
 && -
         \left(\varepsilon_\mu - q_\mu \frac{\varepsilon x}{q x}\right)\left(g_{\beta \nu} -
        \frac{1}{qx} (q_\beta x_\nu + q_\nu x_\beta)\right) q_\alpha
-
         \left(\varepsilon_\nu - q_\nu \frac{\varepsilon x}{q x}\right)\left(g_{\alpha \mu} -
        \frac{1}{qx} (q_\alpha x_\mu + q_\mu x_\alpha)\right) q_\beta
\nonumber \\
 &&+
         \left. \left(\varepsilon_\nu - q_\nu \frac{\varepsilon x}{q.x}\right)\left( g_{\beta \mu} -
        \frac{1}{qx} (q_\beta x_\mu + q_\mu x_\beta)\right) q_\alpha \right]
   \int {\cal D}\alpha_i e^{i (\alpha_{\bar q} + v \alpha_g) qx} {\cal T}_1(\alpha_i)
\nonumber \\
 &&+
        \left[\left(\varepsilon_\alpha - q_\alpha \frac{\varepsilon x}{qx}\right)
        \left(g_{\mu \beta} - \frac{1}{qx}(q_\mu x_\beta + q_\beta x_\mu)\right) \right. q_\nu
\nonumber \\ &&-
         \left(\varepsilon_\alpha - q_\alpha \frac{\varepsilon x}{qx}\right)
        \left(g_{\nu \beta} - \frac{1}{qx}(q_\nu x_\beta + q_\beta x_\nu)\right)  q_\mu
\nonumber \\ && -
         \left(\varepsilon_\beta - q_\beta \frac{\varepsilon x}{qx}\right)
        \left(g_{\mu \alpha} - \frac{1}{qx}(q_\mu x_\alpha + q_\alpha x_\mu)\right) q_\nu
\nonumber \\ &&+
         \left. \left(\varepsilon_\beta - q_\beta \frac{\varepsilon x}{qx}\right)
        \left(g_{\nu \alpha} - \frac{1}{qx}(q_\nu x_\alpha + q_\alpha x_\nu) \right) q_\mu
        \right]      
    \int {\cal D} \alpha_i e^{i (\alpha_{\bar q} + v \alpha_g) qx} {\cal T}_2(\alpha_i)
\nonumber \\
&&+\frac{1}{qx} (q_\mu x_\nu - q_\nu x_\mu)
        (\varepsilon_\alpha q_\beta - \varepsilon_\beta q_\alpha)
    \int {\cal D} \alpha_i e^{i (\alpha_{\bar q} + v \alpha_g) qx} {\cal T}_3(\alpha_i)
\nonumber \\ &&+
        \left. \frac{1}{qx} (q_\alpha x_\beta - q_\beta x_\alpha)
        (\varepsilon_\mu q_\nu - \varepsilon_\nu q_\mu)
    \int {\cal D} \alpha_i e^{i (\alpha_{\bar q} + v \alpha_g) qx} {\cal T}_4(\alpha_i)
                        \right\}~,
\end{eqnarray*}
where $\varphi_\gamma(u)$ is the distribution amplitude of leading twist-2, $\psi^v(u)$,
$\psi^a(u)$, ${\cal A}(\alpha_i)$ and ${\cal V}(\alpha_i)$, are the twist-3 amplitudes, and
$h_\gamma(u)$, $\mathbb{A}(u)$, ${\cal S}(\alpha_i)$, ${\cal{\tilde S}}(\alpha_i)$, ${\cal T}_1(\alpha_i)$, ${\cal T}_2(\alpha_i)$, ${\cal T}_3(\alpha_i)$ 
and ${\cal T}_4(\alpha_i)$ are the
twist-4 photon DAs.
The measure ${\cal D} \alpha_i$ is defined as
\begin{eqnarray*}
\label{nolabel05}
\int {\cal D} \alpha_i = \int_0^1 d \alpha_{\bar q} \int_0^1 d
\alpha_q \int_0^1 d \alpha_g \delta(1-\alpha_{\bar
q}-\alpha_q-\alpha_g)~.\nonumber
\end{eqnarray*}

The expressions of the DAs entering into the above matrix elements are defined as:

\begin{eqnarray}
\varphi_\gamma(u) &=& 6 u \bar u \left( 1 + \varphi_2(\mu)
C_2^{\frac{3}{2}}(u - \bar u) \right),
\nonumber \\
\psi^v(u) &=& 3 \left(3 (2 u - 1)^2 -1 \right)+\frac{3}{64} \left(15
w^V_\gamma - 5 w^A_\gamma\right)
                        \left(3 - 30 (2 u - 1)^2 + 35 (2 u -1)^4
                        \right),
\nonumber \\
\psi^a(u) &=& \left(1- (2 u -1)^2\right)\left(5 (2 u -1)^2 -1\right)
\frac{5}{2}
    \left(1 + \frac{9}{16} w^V_\gamma - \frac{3}{16} w^A_\gamma
    \right),
\nonumber \\
h_\gamma(u) &=& - 10 \left(1 + 2 \kappa^+\right) C_2^{\frac{1}{2}}(u
- \bar u),
\nonumber \\
\mathbb{A}(u) &=& 40 u^2 \bar u^2 \left(3 \kappa - \kappa^+
+1\right)  +
        8 (\zeta_2^+ - 3 \zeta_2) \left[u \bar u (2 + 13 u \bar u) \right.
\nonumber \\ && + \left.
                2 u^3 (10 -15 u + 6 u^2) \ln(u) + 2 \bar u^3 (10 - 15 \bar u + 6 \bar u^2)
        \ln(\bar u) \right],
\nonumber \\
{\cal A}(\alpha_i) &=& 360 \alpha_q \alpha_{\bar q} \alpha_g^2
        \left(1 + w^A_\gamma \frac{1}{2} (7 \alpha_g - 3)\right),
\nonumber \\
{\cal V}(\alpha_i) &=& 540 w^V_\gamma (\alpha_q - \alpha_{\bar q})
\alpha_q \alpha_{\bar q}
                \alpha_g^2,
\nonumber \\
{\cal T}_1(\alpha_i) &=& -120 (3 \zeta_2 + \zeta_2^+)(\alpha_{\bar
q} - \alpha_q)
        \alpha_{\bar q} \alpha_q \alpha_g,
\nonumber \\
{\cal T}_2(\alpha_i) &=& 30 \alpha_g^2 (\alpha_{\bar q} - \alpha_q)
    \left((\kappa - \kappa^+) + (\zeta_1 - \zeta_1^+)(1 - 2\alpha_g) +
    \zeta_2 (3 - 4 \alpha_g)\right),
\nonumber \\
{\cal T}_3(\alpha_i) &=& - 120 (3 \zeta_2 - \zeta_2^+)(\alpha_{\bar
q} -\alpha_q)
        \alpha_{\bar q} \alpha_q \alpha_g,
\nonumber \\
{\cal T}_4(\alpha_i) &=& 30 \alpha_g^2 (\alpha_{\bar q} - \alpha_q)
    \left((\kappa + \kappa^+) + (\zeta_1 + \zeta_1^+)(1 - 2\alpha_g) +
    \zeta_2 (3 - 4 \alpha_g)\right),\nonumber \\
{\cal S}(\alpha_i) &=& 30\alpha_g^2\{(\kappa +
\kappa^+)(1-\alpha_g)+(\zeta_1 + \zeta_1^+)(1 - \alpha_g)(1 -
2\alpha_g)\nonumber +\zeta_2[3 (\alpha_{\bar q} - \alpha_q)^2-\alpha_g(1 - \alpha_g)]\},\nonumber \\
\tilde {\cal S}(\alpha_i) &=&-30\alpha_g^2\{(\kappa -\kappa^+)(1-\alpha_g)+(\zeta_1 - \zeta_1^+)(1 - \alpha_g)(1 -
2\alpha_g)\nonumber +\zeta_2 [3 (\alpha_{\bar q} -\alpha_q)^2-\alpha_g(1 - \alpha_g)]\}.
\end{eqnarray}

Numerical values of parameters used in distribution amplitudes are: $\varphi_2(1~GeV) = 0$, 
$w^V_\gamma = 3.8 \pm 1.8$, $w^A_\gamma = -2.1 \pm 1.0$, $\kappa = 0.2$, $\kappa^+ = 0$, $\zeta_1 = 0.4$, $\zeta_2 = 0.3$.
 
 \section*{Appendix B: The explicit expression of \texorpdfstring{$\Delta^{QCD} $}{}  function}
 In this Appendix, we give the explicit expression for the function $\Delta^{QCD}$ acquired from the light-cone QCD sum rule  in Sect. \ref{formalism}. It is obtained by choosing the $\eslash\pslash$ Lorentz structure as follows:
 \begin{align}
  \Delta^{QCD} &= \frac {1} {9830400\, \pi^8}\Big [ 
   -m_b^{13} \Big( (18 e_d + e_b + 18 e_u) \,I[-7, 5] - 
    15 (e_d + e_u) \,I[-6, 4]\Big) - 
 3 m_b^{11} (25 e_d + e_b + 25 e_u)  I[-6, 5]\Big]\nonumber\\
  &+\frac {m_b^9} {14155776 \, \pi^8} \Big[P_1(12 e_d - 5 e_b + 
        12 e_u ) + 
    1152\, m_s \, \pi^2 P_2 (8 e_d + e_b + 8 e_u) - 
    576 \, m_s\, \pi^2 P_3 (8 e_d + e_b + 
       8 e_u)\Big]\nonumber\\
       & \times I[-5, 3]
       +\frac {m_b^9\, (e_d + e_u)} {655360 \, \pi^8} \Big[ 
   5 m_b^2 I[-5, 4] - 8 I[-5, 5]\Big] 
   +\Big[\frac {m_b^9 \, P_1 (e_d + 
         e_u)} {294912 \, \pi^8} + \frac {m_b^7\, 
      m_s} {24576 \, \pi^6}\Big ( 
      3 \, P_2 \big ((3 e_d + 2 e_b + 
             3 e_u) \nonumber\\
             & \times m_ 0^2 + 
          4 m_b^2 (e_d + e_u) \big) - P_3 \,  
     \big ( m_ 0^2(3 e_d + 2 e_b + 3 e_u) + 6\, 
         m_b^2 (e_d + 
             e_u) \big) \Big) + \frac {m_b^7\, P_2 P_3  } {1536 \, \pi^4} (3 e_d + 2 e_b + 
        3 e_u)
      \Big] I[-4, 2] \nonumber\\
             %
      & +\frac {m_b^7} {7077888\, \pi^8} \Big[ P_1 (33 \
e_d - 5 e_b + 33 e_u) + 
   16  m_s \pi^2 P_2 (23 e_d + 72 e_b + 
      648 e_u) - 576 \, 
  m_s \pi^2 P_3  (9 e_d + e_b + 9 e_u) 
    \Big] \nonumber\\
    &\times I[-4, 3]+\frac {m_b^7} {983040\, \pi^8}\Big[  -15 \, m_b^2 \, (e_u + e_d)\, 
   I[-4, 4] - (9 e_d + e_b - 9 e_u) I[-4, 5]
      \Big]
      +\frac {m_b^5} {147456 \, \pi^6}\Big[-2 m_ 0^2 m_b^2\Big (P_1 \nonumber\\
    & \times (8 e_d - e_b + 8 e_u) + 54 \, 
      m_s P_2 (e_d + e_u)  \Big)   + 
    P_3 \Big (m_s P_1  (8 e_d - e_b + 8 e_u) + 
        36 m_s m_ 0^2 m_b^2 (e_d + e_u) + 
        576 \pi^2 P_2 \nonumber\\
    & \times \big ( 
           m_0^2 (2 e_d + e_b + 2 e_u) - 
            m_b^2 (e_d + e_u)\big) \Big)\Big] I[-3, 1]
            +\frac {m_b^5} {2359296 \, \pi^8}\Big[
   29 e_u P_1 m_b^2 - 
    864 e_b m_ 0^2 m_s \pi^2 P_2 - 
    3456 e_u P_2 \nonumber\\& \times \pi^2 m_s (m_ 0^2 + m_b^2) + 
    288 \pi^2 P_3 \Big (m_s\big (m_ 0^2 (e_b + 
             4 e_u) + 6 e_u m_b^2\big) - 
       16 \pi^2 P_2 (e_b + 4 e_u)\Big) + 
    e_d \Big (29 P_1 m_b^2 - 
        576 \pi^2 \nonumber\\
        & \times \Big (m_b^2 m_s (6 P_2 - 
               3 P_3 ) + 
            m_ 0^2 m_s (6 P_2 - 
               2 P_3) + 
            32 \pi^2 P_2 P_3 \Big)\Big)\Big] I[-3, 2]
+\frac {m_b^5\, (e_d + 
    e_u)} {131072 \pi^8} \Big[  P_1 + 
    64 \pi^2 m_s \nonumber\\
    & \times (2 P_2 - P_3)   \Big] I[-3, 3]
    +\frac {m_b^5} {655360 \, \pi^8} \Big[   
   10 \, m_b^2 \, (e_d + e_u)\, I[-3, 
      4] + (-2 e_d + e_b - 2 e_u)\, I[-3, 5] \Big]
      +\frac {m_b^3} {147456 \pi^6} \Big[  \nonumber\\
      & \times  -2 m_s P_2\Big (P_1 (10 e_d - e_b + 10 e_u) + 
       108 m_ 0^2 m_b^2 (e_d + 
           e_u) \Big) + P_3 \Big ( 
       m_s \big ( 
          P_1 (10 e_d - e_b + 10 e_u)+ 
           72 m_ 0^2 m_b^2  \nonumber\\
          & \times (e_d + e_u) \big) + 
        576 \pi^2P_2 \big (m_ 0^2 (e_d + e_b + 
               e_u) - 2 m_b^2 (e_d + e_u) \big)\Big ) \Big] I[-2, 1]
               +\frac {m_b^3} {7077888 \, \pi^8}\Big[-9 (e_d + 
        e_u)\Big (13 P_1 m_b^2 \nonumber\\
        &- 
       288 \pi^2 (m_b^2 m_s (4 P_2 - 
              2P_3) + 
           m_ 0^2 m_s (3 P_2 - P_3) + 16 \pi^2 P_2 P_3 )\Big) I[-2, 
      2] + \Big ((33 e_d + 5 e_b + 
           33 e_u) P_1 \nonumber\\
           &+ 
       1152 m_s \pi^2 P_2 (e_d - e_b + e_u) - 
       576  m_s \pi^2 P_3 (e_d - e_b + 
           e_u)\Big ) I[-2, 3]\Big]
           -\frac {m_b^3} {3276800 \pi^8}\Big[
   25 (e_d + e_u) I[-2, 4] + (e_d \nonumber\\
   &- 3 e_b + e_u) I[-2, 5]\Big]
   -\frac {m_b} {147456 \, \pi^6}\Big[
   2 m_s P_2 \Big (P_1 (3 \
e_d - e_b + 3 e_u) + 
       54 m_ 0^2 m_b^2 (e_d + 
           e_u)\Big ) +P_3 \Big (- 
          m_s (P_1 \nonumber\\
          & \times (3 e_d - e_d + 3 e_u) + 
           36 m_ 0^2 m_b^2 (e_d + e_u) ) + 
        576 \pi^2 P_2 (-e_b m_ 0^2 + 
            m_b^2 (e_d + e_u))  \Big)  \Big] I[-1, 1]\nonumber\\
            &+\frac {m_b} {2359296 \, \pi^8}\Big[
   288 e_b m_ 0^2 m_s \pi^2 P_2 + 
    m_b^2 (e_d + e_u)  \big (23 P_1 - 
       1152 m_s \pi^2 P_2 \big) + 
    96 \pi^2 P_3 \Big (6 m_b^2 m_s (e_d + e_u) \nonumber\\
    &+ 
        e_b (-m_ 0^2 m_s + 
            16 \pi^2 P_2 )\Big) \Big] I[-1, 2]
            +\frac {m_b} {14155776 \, \pi^8}\Big[  (12 e_d + 12 e_u + 
        5 e_b) P_1 + 
    e_b 576 m_s \pi^2  (-2 P_2 + P_3  ) \Big] \nonumber\\
& \times I[-1, 3]
+\frac {m_b} {9830400 \, \pi^8}\Big[ 
   15 m_b^2 (e_d + e_u) I[-1, 4] + 2 e_b I[-1, 5]\Big]
   +\frac {1} {884736\, m_b\, \pi^6}\Big[
   m_s P_ 1 \Big (6 (e_d + e_u) (10 m_b^2 (2 P_ 2  \nonumber\\
   &- P_ 3)+ 
          m_ 0^2 (3 P_ 2 - P_ 3)) + 
       e_b m_ 0^2 (-3 P_ 2 + P_ 3)\Big) - 
    16 \Big (e_b (27 m_ 0^4 + P_ 1) - 
        3  P_ 2 P_ 3 \pi^2 (e_d + e_u) (9 m_ 0^4 + 2 P_ 1)\Big)\Big] I[0, 
  0]\nonumber\\
  &+\frac {1} {36864\, 
   m_b\, \pi^6}\Big[m_s P_ 1 (2 P_ 2 - P_ 3)(e_b - 8e_d -8 e_u)  + 
    144 m_ 0^2 P_ 2 P_ 3 \pi^2 (4 e_b + e_d + e_u) \Big] I[0, 1]\nonumber\\
    &+\frac {(e_d + e_u)} {2359296 \, 
   m_b\, \pi^8} \Big[-5 m_b^2 P_ 1 I[0, 2] + 
    4 \pi^2 (11 P_ 1 + 960 m_s (2 P_ 2 - P_ 3) ) I[0, 3]\Big] +\frac {2 e_b - 21 (e_d + e_u)} {614400 \, m_b\, \pi^8} I[0, 5]
    \nonumber\\
    &-\frac {m_b} {147456 \, \pi^6}\Big[(e_b - 
       11 e_d -11 e_u) m_s P_ 1 (2 P_ 2 - P_ 3) + 
    576 e_b m_ 0^2 P_ 2 P_ 3  \pi^2\Big] I[1, 0]\nonumber
         \end{align}
         \begin{align}
          &-\frac {m_b^7\, 
  f_ {3\gamma}} {393216 \, \pi^6}\Big[\Big ((e_d + 
          e_u) I_ 1[\mathcal {A}] + (e_d - e_u) I_ 1[
          \mathcal {V}]\Big) \Big (m_b^2 I[-5, 4] - 
       3 I[-4, 4]\Big)\Big]
       +\frac {m_b^5\, 
  f_ {3\gamma}} {4718592 \, \pi^6}\Big[\Big (((7 e_d - 4 e_s - 
             21 e_u) P_1\nonumber\\
             &- 
          576 (e_d - 3 e_u) m_s \pi^2 (2 P_ 2 - 
             P_ 3) ) I_ 1[\mathcal {A}] + (-3 e_d P_ 1 + 
          11 e_u P_ 1 + 
          576 \pi^2 (e_d - e_u) m_s (2 P_ 2 - 
             P_ 3)) I_ 1[\mathcal {V}]\Big) I[-3, 2] \nonumber\\
          & - 
    36 \Big((e_d + e_u) I_ 1[\mathcal {A}] + (e_d - 
          e_u) I_ 1[\mathcal {V}]\Big) I[-3, 4]\Big]
         +\frac {m_b^3\, 
  f_ {3\gamma}} {4718592 \, \pi^6} \Big[\Big (192 \pi^2 (5 e_d - 
           3 e_u) \big (m_ 0^2 m_s (3 P_ 2 - P_ 3) + 
          16 P_ 2 P_ 3 \pi^2\big)\nonumber\\
          & \times I[-2,1] + \big ((-21 e_d + 4 e_s + 35 e_u) P_ 1 + 
          576 \pi^2 (3 e_d - 5 e_u) m_s (2 P_ 2 - P_ 3) \big) I[-2, 
          2]\Big) I_ 1[\mathcal {A}] + (192 \pi^2 (e_d - 
            e_u) \big (m_ 0^2 m_s \nonumber\\
            & \times (3 P_ 2 - P_ 3) + 
           16\pi^2 P_ 2 P_ 3 \big) I[-2, 
          1] - \big ((e_d + 15 e_u) P_ 1 + 
           576 \pi^2 (e_d - e_u) m_s (2 P_ 2 - P_ 3)\big) I[-2, 
          2]) I_1[\mathcal {V}]\Big]\nonumber\\
          &+\frac {f_ {3\gamma}} {24576 \, 
   m_b \pi^4}\Big (m_ 0^2 m_s (3 P_ 2 - P_ 3) + 
    16 \pi^2 P_ 2 P_ 3 \Big) \Big ((e_d + 
      e_u) I_ 1[\mathcal {A}] + (e_d - e_u) I_ 1[\mathcal {V}]\Big) I[
  0, 1],  
  \label{app}
         \end{align}
where $P_1 =\langle g_s^2 G^2\rangle$ is gluon condensate, $P_2 =\langle \bar q q \rangle$ stands for u/d quark condensate, and $P_3 = \langle \bar s s \rangle$ represents s-quark condensate. It should be noted that in Eq. (\ref{app}), for the sake of brevity, we only give  expressions that make substantial contributions to the numerical value of the magnetic moment and do not present much higher-dimensional contributions, even though they are considered in numerical calculations.

The functions~$I[n,m]$ and  $I_1[\mathcal{A}]$  
are
defined as:
\begin{align}
 I[n,m]&= \int_{m_b^2}^{s_0} ds \int_{m_b^2}^s dl~ e^{-s/M^2}~\frac{(s-l)^m}{l^n},\nonumber\\
I_1[\mathcal{F}]&=\int D_{\alpha_i} \int_0^1 dv~ \mathcal{F}(\alpha_{\bar q},\alpha_q,\alpha_g)
 \delta(\alpha_{\bar q}+ v \alpha_g-u_0),
 \end{align}
 where $\mathcal{F}$ stands for the corresponding photon DAs.

\end{widetext}

\bibliography{XibMM}
\end{document}